\documentclass{IEEEtran}
\usepackage{cite}
\usepackage{amsmath,amssymb,amsfonts}
\usepackage{algorithmic}
\usepackage{graphicx}
\usepackage{textcomp}
\usepackage{subfigure}
\def\BibTeX{{\rm B\kern-.05em{\sc i\kern-.025em b}\kern-.08em
    T\kern-.1667em\lower.7ex\hbox{E}\kern-.125emX}}
\begin{document}
\title{A Study Of a Wide-Angle Scanning Phased Array Based On a High-Impedance Surface Ground Plane}
\author{Tian Lan, Qiu-Cui Li, Yu-Shen Dou, Xun-Ya Jiang
\thanks{This work was supported by the National Natural Science Foundation of China under Grant 11334015, and the National Key Research and Development Project of China under Grant 2016YFA0301103, 2018YFA0306201.}
\thanks{These authors are with the Department of Light Source \& Illuminating Engineering, Fudan University, Shanghai 200344, China (e-mail: 	
	jiangxunya@fudan.edu.cn).}
}

\maketitle

\begin{abstract}
This paper presents a two-dimensional infinite dipole array system with a mushroom-like high-impedance surface (HIS) ground plane with wide-angle scanning capability in the E-plane. The unit cell of the proposed antenna array consists of a dipole antenna and a four-by-four HIS ground. The simulation results show that the proposed antenna array can achieve a wide scanning angle of up to 65$^{\circ}$ in the E-plane with an excellent impedance match and a small $S11$. Floquet mode analysis is utilized to analyze the active impedance and the reflection coefficient. Good agreement is obtained between the theoretical results and the simulations. Using numerical and theoretical analyses, we reveal the mechanism of such excellent wide scanning properties. For the range of small scanning angles, these excellent properties result mainly from the special reflection phase of the HIS ground, which can cause the mutual coupling between the elements of the real array to be compensated by the mutual coupling effect between the real array and the mirror array. For the range of large scanning angles, since the surface wave (SW) mode could be resonantly excited by a high-order Floquet mode $\textrm{TM}_{-1,0}$ from the array and since the SW mode could be converted into a leaky wave (LW) mode by the scattering of the array, the radiation field from the LW mode is nearly in phase with the direct radiating field from the array. Therefore, with help from the special reflection phase of the HIS and the designed LW mode of the HIS ground, the antenna array with an HIS ground can achieve wide-angle scanning performance. 
\end{abstract}

\begin{IEEEkeywords}
High impedance surfaces (HISs), phased arrays, wide-angle scanning, surface waves(SWs).
\end{IEEEkeywords}

\section{Introduction}
\label{sec:introduction}
Generally, the main beam of a planar phased array cannot effectively scan to large angles due to the mutual coupling among the antenna elements and the excited surface waves (SWs), which can cause the reflection coefficient $S11$ to increase rapidly\cite{b11}. Several different approaches have been applied to improve the radiation performance of planar phased arrays, such as a subarray technique for suppressing SWs\cite{b1}, inhomogeneous substrates\cite{b2}, reduced surface wave (RSW) antenna elements\cite{b3}, and defected ground structures\cite{b4}.

In recent years, there has been increasing interest in utilizing high-impedance surface (HIS)\cite{b5} structures in array design. Because of their unique reflection phase and bandgap characteristics, HISs provide a new degree of freedom in antenna design; for example, HISs are widely used as the ground planes of arrays to suppress SW generation using the HIS gap\cite{b6,Donzelli2007,Li2007}. They can also be placed between array elements\cite{b7,Azarbar2011} to reduce the mutual coupling between those elements to extend the scanning range of the beam. 

Recently, researchers have explored whether the SW modes supported by HISs can help to improve certain aspects of the radiation performance of antennas or antenna arrays. In\cite{b8}, it is shown that the TE SW is resonantly excited and the edge radiation is favorable for broadening the bandwidth and maintaining the radiating pattern in the bandwidth. Li et al.\cite{b9} proposed that one dipole antenna and two parasitic elements be placed in close proximity to a finite HIS ground. Using the advantage of TE SW propagation on an HIS and the HIS edge radiation, a wide beam tilting toward the endfire direction is achieved. Then, Li et al.\cite{b10} designed an HIS-based linear array with eight dipoles whereby the HIS edge radiation of the SW supported by the HIS is also utilized to achieve wide-angle scanning in the H-plane. Thus, these works demonstrate that an HIS SW that causes HIS edge radiation can improve the radiation performance for single elements or small arrays on an HIS. However, this method cannot be applied to a large array on an HIS since the importance of edge radiation will be significantly reduced with an increasing number of array elements, and scan blindness may occur because the SW can absorb large amounts of radiating energy. For large antenna arrays and infinite arrays, is it possible to find a design whereby the SW mode supported by the HIS ground can improve the wide-angle scanning performance? To the best of our knowledge, there is no relevant research on this topic.

\begin{figure*}[!t]
	\centerline{\includegraphics[width=1.5\columnwidth]{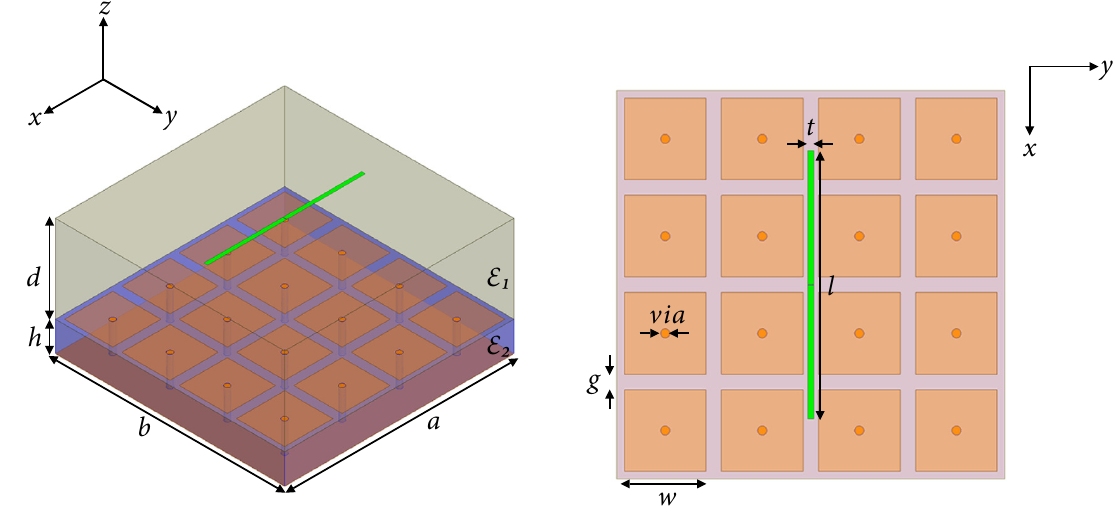}}
	\caption{Structure of the dipole array on the HIS ground plane. The top figure gives a view of a unit cell of the infinite phased dipole array printed on the HIS ground plane. The bottom figure shows 4-by-4 unit cells of the HIS ground plane. Some key parameters are as follows: the square patch size $w=3.15$~mm, the gap between patches $g=0.6$~mm, the via size $via=0.36$~mm, the dipole size $l=10.32$~mm (length) and $t=0.06$~mm (width), the substrate thickness $h=1.95$~mm and $d=5.7$~mm, the lattice constant is $a=b=30$~mm, and the substrate permittivity $\varepsilon_{1}=$2.55 and $\varepsilon_{2}$=4.4.}
	\label{mushroom_dipole_array}
\end{figure*}
In this paper, we will design a two-dimensional infinite dipole array system with a mushroom-like HIS ground plane. With the unique reflection phase characteristics of the HIS ground plane, this array can achieve a wide scanning angle of up to 65$^{\circ}$ in the E-plane with a small $S11$. Then, we will analyze the relationship between the reflection phase of the HIS and the active impedance of the array by Floquet mode analysis, demonstrate that the reflection phase of HIS is a parameter that is critical to the antenna's radiation performance, and reveal the mechanisms behind our design. We find that there are two mechanisms supporting the wide-angle performance in such infinite arrays: (i) The coupling effect between real antenna elements and the mirror antenna elements with an HIS as the ground can cancel the mutual coupling between the real antenna elements. This canceling ensures the very good radiation performance for a small scanning angle range {0$^{\circ}$ - 20$^{\circ}$}. (ii) For large scanning angles of {20$^{\circ}$ - 65$^{\circ}$}, the downward high-order Floquet radiating field from the antenna array can excite the SW mode of the HIS, and with periodic scattering of the antenna array, such an SW mode can be transformed into a leaky wave (LW) mode\cite{Tamir1973}. Due to the specially designed reflection phase of the HIS, the radiation from the LW mode can be coherently added to the direct upward radiating field from the antenna array in a wide angle range. The imaginary part of the active impedance is thereby maintained at a small value, while the real part remains almost constant over a wide scanning angle range.

This paper is structured as follows. An infinite two-dimensional dipole array with a mushroom-like HIS ground plane is proposed, and the simulation results of the active impedance and reflection coefficient $S11$ of our design are presented in Sec \ref{design}. Then, we use Floquet mode analysis to calculate the active impedance and the reflection coefficient $S11$ of the system and show the relationship between the reflection phase of the HIS and the active impedance of the array in Sec \ref{FloquetMode}. The mechanisms of the excellent performance of our design are analyzed in detail in Sec \ref{mirror} and Sec \ref{leakywave}. Finally, we conclude our paper in Sec \ref{conclusion}.

\section{Dipole Array Design Based on HIS Ground Plane and HFSS Simulations}
\label{design}

In general, to ensure excellent radiation performance, we hope that the field reflected by the ground plane will be in-phase with the direct radiation field of the antenna array. However, for traditional design with the PEC as ground, we can only guarantee the "in-phase" property for one angle (e.g., the zero scanning angle) since the reflection phase is a constant. The phase difference between the reflected field and the direct radiating field increases when the scanning angle becomes larger. Because the reflection phase of HIS varies with the incident angle, it is possible to achieve the "in phase" property within a certain range of scanning angle if we design a special HIS as ground. In addition, because of its complex and unique reflection phase, the SW mode of HIS ground can be very different from the traditional SW mode of PEC which will weaken the radiation performance of arrays in general. We will show that the SW mode of the HIS can greatly improve the radiation performance of antenna arrays under certain design.

After optimization of the parameters of dipole antenna, the HIS ground and the dielectric substrate between them, we design 
a two-dimensional infinite dipole array, and the simulation model of the unit cell of our design is shown in Fig. \ref{mushroom_dipole_array}. This unit cell consists of a dipole antenna printed on substrate back by the four-by-four HIS ground plane. The lattice constant is $a=b=\lambda/2=30$~mm, where $\lambda$ denotes the wavelength in free space at the operation frequency 10 GHz. The length and width of the infinitely thin dipole are $l= 10.32$~mm and $t=0.06$~mm respectively. The region between dipole and HIS is filled with dielectric substrate with the thickness $d = 5.7$~mm and the permittivity $\varepsilon_{1}=2.55$. For HIS design, an infinitely thin square patch with a side length of $w=3.15$~mm is printed on top of a grounded substrate with a dielectric constant of $\varepsilon_{2}=4.4$ and a thickness of $h=1.95$~mm. The length of the gap between adjacent patches is $g=0.6$~mm. Vertical conducting paths with a diameter of $via=0.36$~mm are used to connect the upper patches to the ground plane. 

The infinite array performance was analyzed based on this unit cell using a commercial full-wave EM simulation software High Frequency Structure Simulation (HFSS) which applies Floquet's theorem of periodic boundaries. While this method accounts for the mutual coupling between the array elements, it does not include the effect of edge elements in the case of finite arrays. In the simulation setup, periodic boundaries are used at the sides of the unit cell of antenna array in both $x$ and $y$ directions, and a Floquet port terminates the setup from the top. The radiating modes from the structure surface propagate
within air, filled between the unit cell surface and Floquet port, and are absorbed from the top. The dipole is fed at the center by a lumped port with a port impedance of 16 ohms so as to match the input impedance at broadside. Next, the active input impedance, the magnitude of the reflection coefficient $S11$ versus the scanning angle and the scanning performance will be calculated by ANSYS HFSS simulations.

\begin{figure}[!t]
	\centerline{\includegraphics[width=\columnwidth]{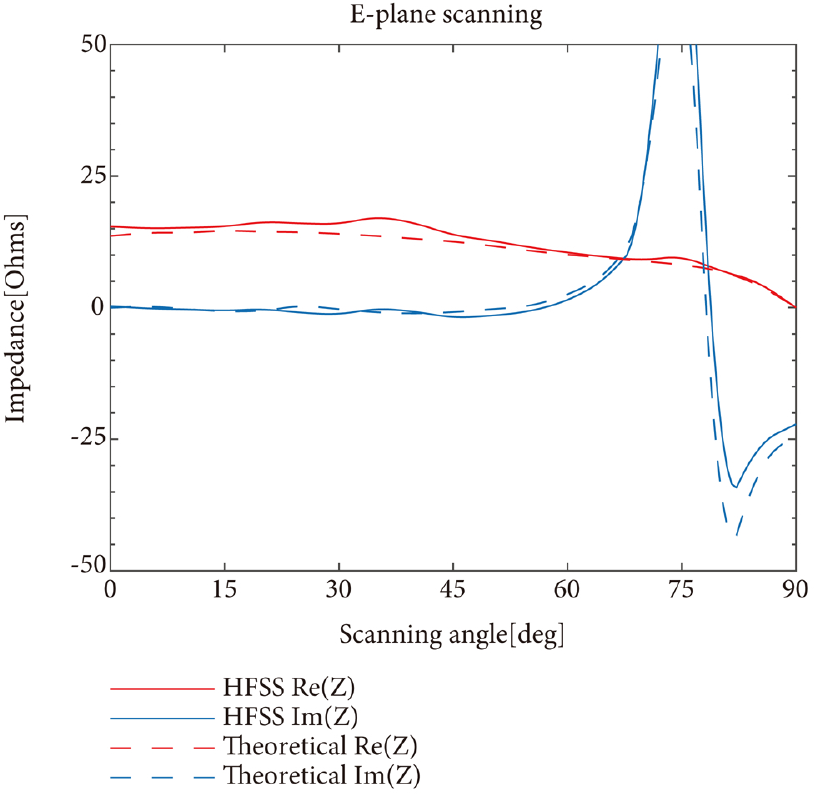}}
	\caption{Active input impedance of the dipole array on HIS ground plane shown in Fig. \ref{mushroom_dipole_array} during an E-plane scan, where solid lines are obtained from HFSS and dashed lines are calculated from Eq. \ref{eq:active_impedance}.}
	\label{active_impe}
\end{figure}

Fig. \ref{active_impe} shows the active impedance variations during an E-plane scan, where the solid lines are obtained from HFSS simulations, while the dashed lines will be explained in the next section. It can be seen that the imaginary part of active impedance is maintained at a small value, while the real part remains almost constant within the scanning angle range of $0^{\circ}-65^{\circ}$, which indicates that the array exhibits excellent impedance-matching performance.

\begin{figure}[!t]
	\centerline{\includegraphics[width=\columnwidth]{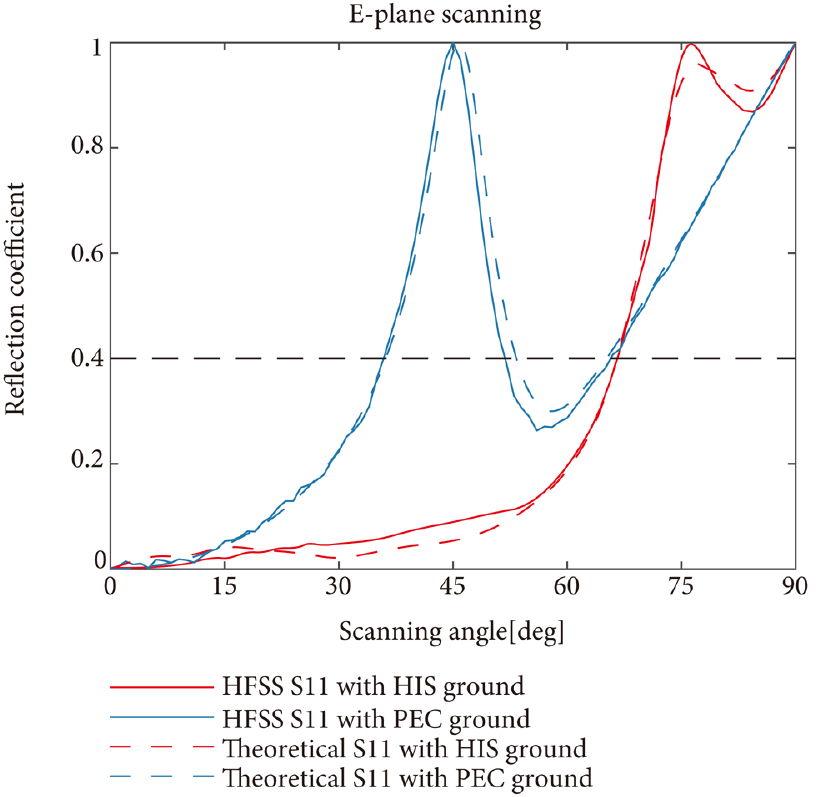}}
	\caption{Comparison of the reflection coefficient in the E-plane scan for the dipole array on the HIS ground plane and PEC ground plane, where the solid lines are obtained from HFSS and the dashed lines are calculated from Eq. \ref{eq:active_impedance}. For the case of the PEC ground plane, not only is the HIS ground replaced by the PEC, but also the dipole size is tuned to have a resonance at broadside.}\label{hiss11}
\end{figure}

\begin{figure}[!t]
	\centerline{\includegraphics[width=\columnwidth]{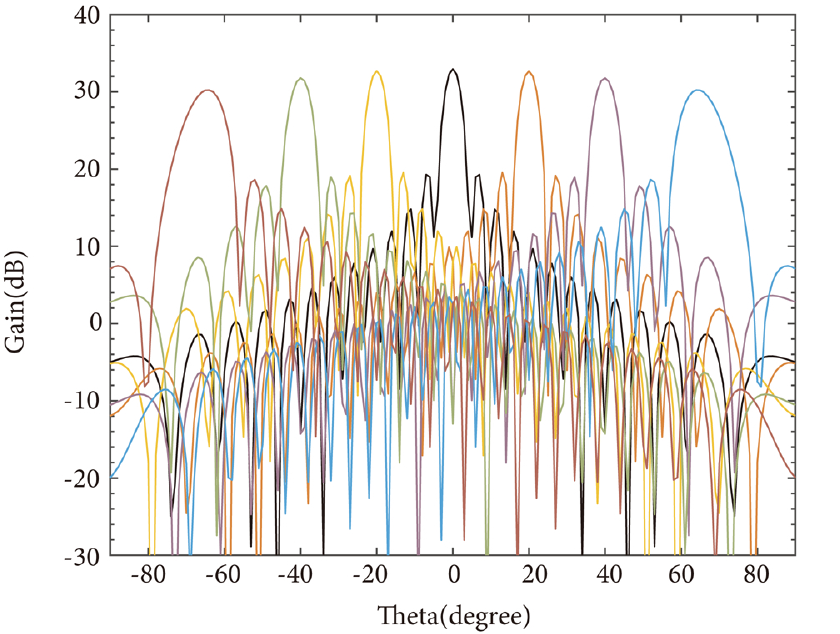}}
	\caption{Simulated pattern scanning characteristics in the E-plane at 10 GHz with the main beam pointing direction of $\theta = 0^\circ, \pm 20^\circ, \pm 40^\circ, \pm 65^\circ$ , respectively.}\label{radPattern}
\end{figure}

We calculate the magnitude of the reflection coefficient $S11$ of the HIS-ground-plane-based array during an E-plane scan and compare it with that of an array with a PEC ground plane, as shown in Fig. \ref{hiss11}, where the red solid line is the case of the HIS ground and the blue solid line is the case of the PEC ground plane. The comparison reveals that the impedance-matching performance of the array with the HIS ground plane is significantly better than that of the array with the PEC ground plane. From Fig. \ref{hiss11}, we can see that the array can achieve a wide scanning angle of up to 65$^\circ$ with $S11<0.4$. In addition, for the case of the HIS ground plane,  the simulated scan performance of our array in the E-plane at 10 GHz is shown in Fig. \ref{radPattern}. We can see that the main beam of our array can scan from -65$^\circ$  to +65$^\circ$ in the E-plane with a gain fluctuation less than 3 dB and a maximum sidelobe level (SLL) less than -10dB. The radiation patterns corresponding to the main beam toward 0$^\circ$, $\pm$20$^\circ$,
$\pm$40$^\circ$, $\pm$65$^\circ$ are particularly plotted in Fig. \ref{radPattern}. By contrast, for the case of the PEC ground, the array can only scan its main beam to 35$^\circ$ at the same standard, and scan blindness appears at 45$^\circ$ since the SW mode is excited. As a result, the proposed array can scan its main beam over the range from -65$^\circ$ to +65$^\circ$.In the next section, we will analyze why the proposed system can achieve wide-angle scanning.

\section{Discussion}
In the previous section, we introduced the design of a infinite dipole array that can achieve wide-angle scanning in the E-plane . In this section, we will analyze the relationship between the HIS reflection phase and the active impedance of the array via Floquet mode analysis to show that the reflection phase of HIS is a parameter that is critical to the antenna's radiation performance, thereby revealing the mechanisms behind our design.

\subsection{Floquet Mode Analysis}
\label{FloquetMode}
In this subsection, Floquet mode analysis is used to calculate the active input impedance and the magnitude of the reflection coefficient $S11$ versus the scanning angle. Then, we compare the theoretical results from the Floquet mode analysis with those of the HFSS simulations. Additionally, the effect of the HIS reflection phase is clearly shown in the analysis.

According to Floquet mode analysis\cite{bhattacharyya2006phased}, the active input impedance $Z^{\textrm{FL}}$ of an infinite antenna array with a general ground can be obtained by

\begin{equation}
\begin{split}
Z^{\textrm{FL}}(k_{x0},k_{y0}) = &\frac{4}{ab} \frac{l^2}{\pi^2}\sum_{m=-\infty}^{\infty}\sum_{n=-\infty}^{\infty}\left ( \frac{k^2_{ymn}}{y_{mn}^\textrm{TE}} + \frac{k^2_{xmn}}{y_{mn}^\textrm{TM}} \right ) \\
&\left [ \frac{\cos(k_{xmn}l/2)}{1-(k_{xmn}l/\pi)^2} \right ]  \frac{\sin c(k_{ymn}t/2)}{k_0^2-k_{zmn}^{+2}} \\
\end{split}
\label{eq:active_impedance}
\end{equation}
with
\begin{align}
\label{eq:ymnTE}
& y_{mn}^{\textrm{TE}} = Y_{mn}^{\textrm{TE+}} - jY_{mn}^{\textrm{TE-}}\cot(k_{zmn}^-h - \theta_{mn}^{\textrm{TE}}/2) \\
&Y_{mn}^{\textrm{TE+}} = \frac{\omega\epsilon_0}{k_{zmn}^+}\;\;\;\;\;\;\;Y_{mn}^{\textrm{TM-}} =\frac{\omega\epsilon_0\epsilon_r}{k_{zmn}^-}
\end{align}
\begin{align}
\label{eq:ymnTM}
& y_{mn}^{\textrm{TM}} = Y_{mn}^{\textrm{TM+}} - jY_{mn}^{\textrm{TM-}}\cot(k_{zmn}^-h - \theta_{mn}^{\textrm{TM}}/2) \\
&Y_{mn}^{\textrm{TM+}} = \frac{k_{zmn}^+}{\omega \mu_0}\;\;\;\;\;\;\;Y_{mn}^{\textrm{TE-}} =\frac{k_{zmn}^-}{\omega \mu_0}
\end{align}
\begin{equation}
\label{eq:phase_const}
k_{xmn} = k_{x0} + \frac{2m\pi}{a}\; \; \; \; \; \; \; \; \; \;
k_{ymn} = k_{y0} + \frac{2n\pi} {b}
\end{equation}
\begin{equation}
k_{zmn} = \sqrt {k^{2} - k_{xmn}^{2} - k_{ymn}^2}
\end{equation}
where $k$ is the wavenumber in a medium or in free space, $\theta_{mn}^{\textrm{TE/TM}}$ are the reflection phases of the Floquet modes reflected by a general ground, e.g., the HIS in our design. In Eq. \ref{eq:phase_const}, $k_{x0}$ and $k_{y0}$ are phase progression factors related to the intended direction of radiation. If $(\theta,\phi)$ are angles in spherical coordinate system related to the intended direction of radiation, then
\begin{equation}
k_{x0} = k_{0} \sin \theta \cos \phi \; \; \; \; \; \; \; \; \; \;
k_{y0} = k_{0} \sin \theta \sin \phi
\end{equation}
We note that the term in Eq. \ref{eq:active_impedance} with the reflection phase of the HIS ground shows the contribution of reflected waves to the active impedance.

From Eq.\ref{eq:active_impedance}, we can calculate the active impedance and reflection coefficient $S11$ and compare the results with those of the HFSS simulations. If the results fit very well, then we have confidence that our analysis is correct. However, in order to calculate the active impedance, we must first obtain the reflection phases $\theta_{mn}^{\textrm{TE/TM}}$ of the HIS.

We calculate the reflection phases of different orders of Floquet modes using EastWave commercial software based on the finite-difference time-domain (FDTD) method. We can then bring the reflection phases $\theta_{mn}^{\textrm{TE/TM}}$ into Eq.\ref{eq:active_impedance} and obtain the contribution to the active impedance from all Floquet modes, as shown by the red and blue dashed lines in Fig. \ref{active_impe}. Meanwhile, we can calculate the reflection coefficient $S11$ by Eq. \ref{eq:active_impedance}. We find that the theoretical results are in good agreement with the simulation results. Moreover, from the calculated results, we find that the most important Floquet modes for our antenna array which can affect the impedance and $S11$ are the $\textrm{TM}_{0,0}$ mode and the $\textrm{TM}_{-1,0}$ mode. This finding is easy to understand for two reasons. The first is that we consider only E-plane scans in this work, so the $\textrm{TM}$ modes dominate the far-field radiation. The second is that except for the $\textrm{TM}_{0,0}$ mode and the $\textrm{TM}_{-1,0}$ mode, all $\textrm{TM}$ modes are evanescent waves in all scanning angle ranges, and they contribute only small perturbations of the active impedance and $S11$.
\begin{figure}[!t]
	\centerline{\includegraphics[width=\columnwidth]{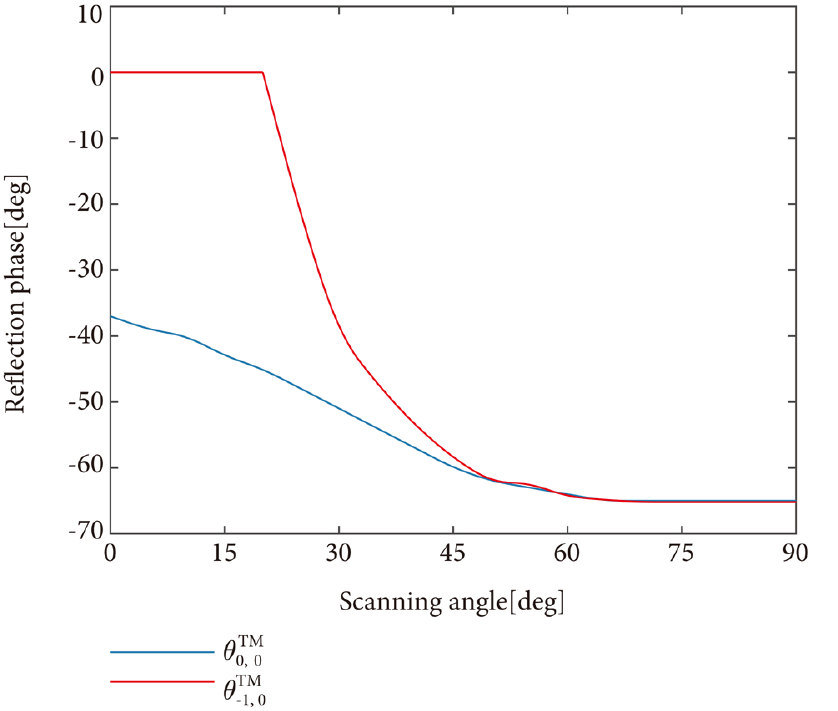}}
	\caption{Reflection phases of the Floquet modes reflected at the HIS versus the scanning angle, where the blue (red) solid line is the reflection phase of the $m = 0(m=-1)$-order Floquet mode, and the remaining reflection phases are approximately equal to zero.}
	\label{theta00&-10reflectionphase}
\end{figure}
\begin{figure}[!t]
	\centerline{\includegraphics[width=\columnwidth]{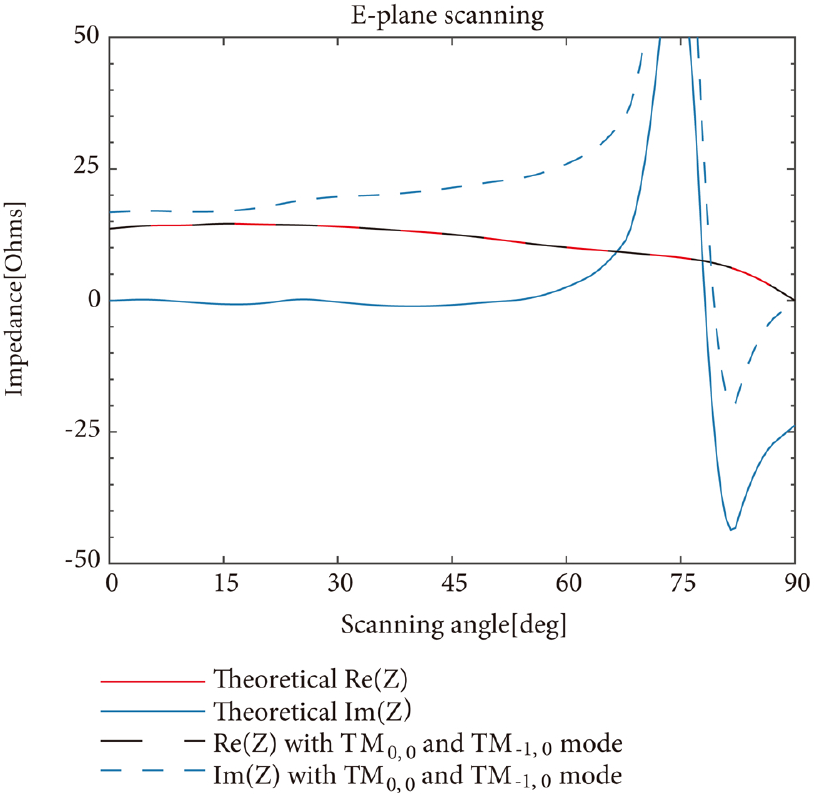}}
	\caption{Contribution of $\textrm{TM}_{0,0}$ and $\textrm{TM}_{-1,0}$ mode to the active impedance of the array, where the dashed lines are obtained with only the contributions of the two modes and the solid lines are the results of the dashed lines in Fig. \ref{active_impe}}
	\label{twomode}
\end{figure}

The calculated results of the reflection phases of the $\textrm{TM}_{0,0}$ mode and the $\textrm{TM}_{-1,0}$ mode are shown in Fig. \ref{theta00&-10reflectionphase}. We can see that the reflection of the $\textrm{TM}_{0,0}$ mode decays almost linearly with the scanning angle within the range of $0^\circ-20^\circ$, which is very important for the excellent radiating properties for small scanning angles. For the $\textrm{TM}_{-1,0}$ mode, its field is an evanescent wave so that the reflection phase is zero when the scanning angle is smaller than 20$^\circ$. Once the scanning angle is larger than 20$^\circ$, the field of the $\textrm{TM}_{-1,0}$ mode propagates in the substrate material and the reflection phase of this mode is no longer zero. With the reflection phases, the contributions from both the $\textrm{TM}_{0,0}$ mode and the $\textrm{TM}_{-1,0}$ mode on the active impedance are shown by the dashed lines in Fig. \ref{twomode}. The real part of the active impedance obtained from Eq.\ref{eq:active_impedance} with the contributions of only these two modes is in good agreement with the result obtained from all modes, while the changing trend of the imaginary part of the impedance is essentially the same as that with the contributions of all modes. Therefore, there must be some basic mechanisms which support such good agreement. In the next two subsections, the mechanisms of the excellent performance of the array with the HIS ground plane will be analyzed in detail.

\subsection{Effect of the HIS Ground Plane in a Range of Small Scanning Angles}
\label{mirror}

In this subsection, we demonstrate how the HIS ground plane improves the scanning performance of the antenna array over the range of small scanning angles $0^\circ-20^\circ$ considering the special reflection phase of the HIS. For the small scanning angles, the dominant mode is the $\textrm{TM}_{0,0}$ mode because the field of the $\textrm{TM}_{-1,0}$ mode is still an evanescent wave.

\begin{figure}[!t]
	\centerline{\includegraphics[width=\columnwidth]{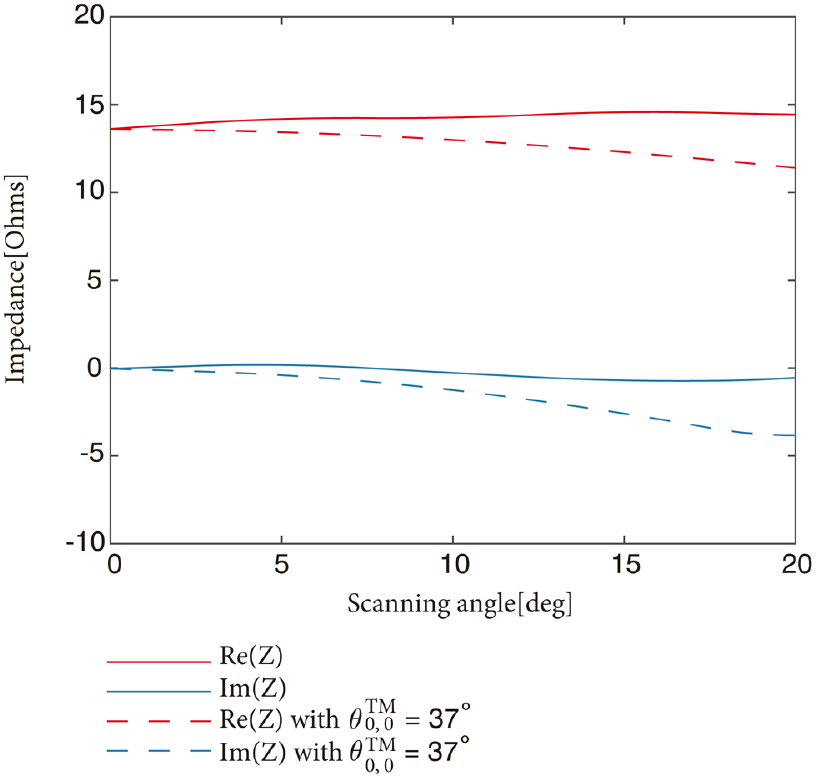}}
	\caption{Effect of $\theta_{0,0}^{\textrm{TM}}$ on active impedance in a small scanning angle range. The dashed line is the case where the reflection phase of $\theta_{0,0}^{\textrm{TM}}$ is a constant, while the solid lines are the same as the results of the dashed lines in Fig. \ref{active_impe}.}
	\label{theta00}
\end{figure}

First, we make a naive assumption that the reflection phase of the $\textrm{TM}_{0,0}$ mode from the HIS ground is constant, i.e., it is maintained at the value of zero scanning angle $- 37^{\circ}$ for any scanning angle, while the reflection phases $\theta_{mn}^{\textrm{TE/TM}}$ of the other modes still change in their original manner. Then, we can calculate the active impedance versus the scanning angle using Eq.\ref{eq:active_impedance} and the result is shown by the dashed lines in Fig. \ref{theta00}. Compared with the original calculated results shown by the solid line in Fig. \ref{theta00}, we can see that if the HIS reflection phase of $\textrm{TM}_{0,0}$ mode were a constant similar to PEC, the original excellent properties such as the almost-constant real part and the nearly-zero imaginary part at scanning angles less than 20$^{\circ}$, would be destroyed. Clearly, the only explanation for such destruction is that the changing HIS reflection phase versus the scanning angle shown by the blue line in Fig. \ref{theta00&-10reflectionphase} is very critical for small scanning angles.

\begin{figure}[!t]
	\centerline{\includegraphics[width=\columnwidth]{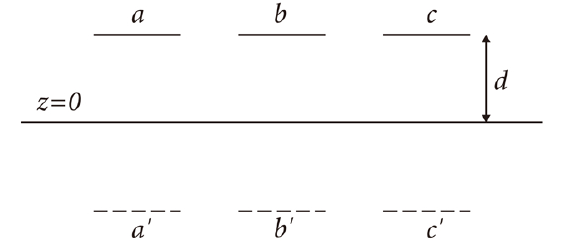}}
	\caption{Simplified structure of the dipole array, where $a$, $b$, and $c$ are real dipoles; $a'$, $b'$, and $c'$ are image dipoles, and an HIS ground plane is placed at $z=0$.}\label{simple_dipole_array}
\end{figure}

From the view of an image antenna, we can more clearly see the effect of the reflection phase of an HIS. We emphasize that there are two different approaches to study the physical effects of the reflected field from the ground. One approach is to study the reflected field directly as we did before. The other approach is to introduce the image antenna (or image antenna array) whose radiating field is substituted by the reflected field with the exact same phase and magnitude. Hence, the effect of the reflected field on the real antenna array could be viewed as the coupling between the real antenna array and the image antenna array. To clearly show the coupling effects on the active impedance, we simplify the infinite two-dimensional array to the model shown in Fig. \ref{simple_dipole_array}, where $a$, $b$, and $c$ are real dipoles on an infinitely large ground with reflection phase $\theta_r$. Without the ground plane, the impedance of antenna b could be obtained by:
\begin{equation}
\label{eq:mutal impe_without}
\begin{split}
Z_{b,in}^0 = & Z_b + Z_{ab}{\rm e}^{j\theta_{ab}} + Z_{cb}{\rm e}^{j\theta_{cb}} \\
\end{split}
\end{equation}
where $Z_b$ is the self-impedance, $Z_{ij}$ is the mutual impedance between elements $i$ and $j$, and $\theta_{ij}$ is the input current phase difference between elements $i$ and $j$. It is well known that the $Z_{b,in}^0$ changes with the scanning angle since the input current phase difference $\theta_{ij}$ changes with the scanning angle.

With a ground, image dipoles $a'$, $b'$, and $c'$ should be introduced and the active impedance of element $b$ is:
\begin{equation}
\label{eq:mutal impe}
\begin{split}
Z_{b,in} = & Z_b + Z_{ab}{\rm e}^{j\theta_{ab}} + Z_{cb}{\rm e}^{j\theta_{cb}} \\
&+ Z_{a'b}{\rm e}^{j(\theta_{ab}+\theta_r)} + Z_{b'b}{\rm e}^{j\theta_{r}} + Z_{c'b}{\rm e}^{j(\theta_{cb}+\theta_r)} \\
\end{split}
\end{equation}
From eq. \ref{eq:mutal impe}, we can see that the active impedance of antenna b varies with the scanning angle if the reflection phase of the surface is constant. However, for an HIS, the reflection phase decreases almost linearly with the scanning angle as shown by the blue line in Fig. \ref{theta00&-10reflectionphase}. With increasing scanning angle, the change caused by $Z_{ab}{\rm e}^{j\theta_{ab}} + Z_{cb}{\rm e}^{j\theta_{cb}}$ could be canceled by the change caused by $Z_{a'b}{\rm e}^{j(\theta_{ab}+\theta_r)} + Z_{b'b}{\rm e}^{j\theta_{r}} + Z_{c'b}{\rm e}^{j(\theta_{cb}+\theta_r)}$. In other words, the mutual coupling effect between elements of a real array at small scanning angles can be compensated by the mutual coupling effect from the mirror array, thereby greatly improving the radiation performance of the antenna array with an HIS ground.

\subsection{Effect of the HIS Ground Plane Over a Range of Large Scanning Angles}
\label{leakywave}
In the previous subsection, the improving of radiation efficiency in small scanning angles is explained. In this subsection, we will demonstrate the new mechanism of the HIS ground plane in improving the scanning performance of the antenna array over a range of large scanning angles, and reveal the effect of the LW mode.

\begin{figure}[!t]
	\centerline{\includegraphics[width=\columnwidth]{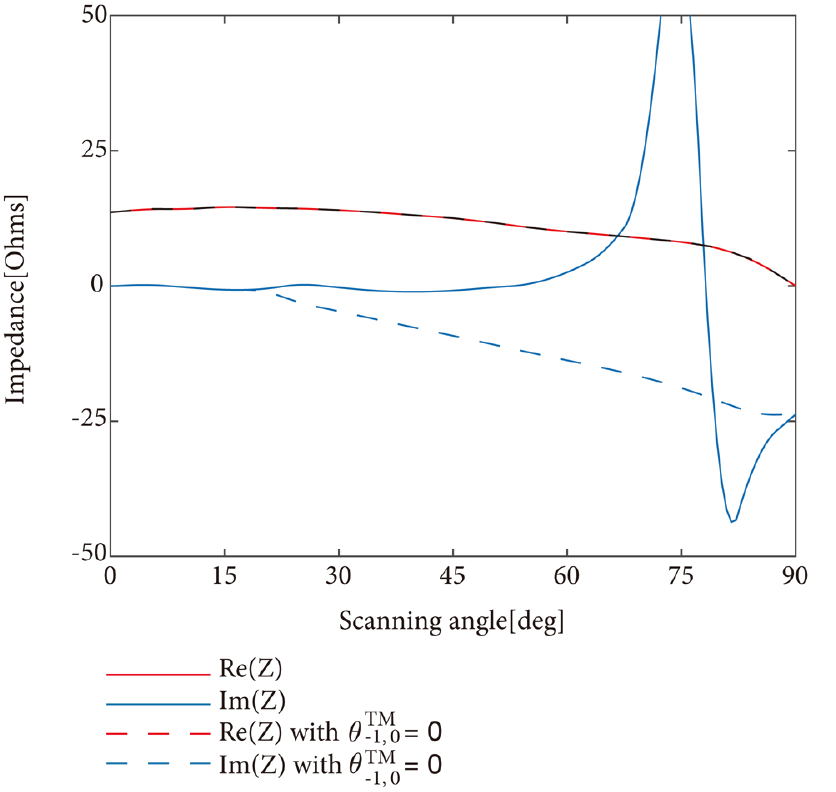}}
	\caption{Effect of $\theta_{-1,0}^{\textrm{TM}}$ on the active impedance over a range of large scanning angle. The dashed line is the case where the reflection phase of $\theta_{-1,0}^{\rm TM}$ is zero, while the solid lines are the same as the results of the dashed lines in Fig. \ref{active_impe}.}
	\label{theta-10effect}
\end{figure}

First, we detect the effect of the reflection phase of the HIS on the $\textrm{TM}_{-1,0}$ Floquet mode, shown by the red line in Fig. \ref{theta00&-10reflectionphase}. Similarly, at first we assume that the reflection phase of the $\theta_{-1,0}^{\textrm{TM}}$ of $\textrm{TM}_{-1,0}$ mode from the HIS ground always is zero for any scanning angle, while the reflection phases $\theta_{mn}^{\textrm{TE/TM}}$ of all other modes still change in their original manner. We can then calculate the active impedance versus the scanning angle by Eq.\ref{eq:active_impedance}. The results are shown by the dashed lines in Fig. \ref{theta-10effect}. We can see that when the scanning angle exceeds $20^{\circ}$, the imaginary part of the active impedance begins to deviate from the original value and for larger scanning angles the deviating values become larger, which indicates that the coupling effect between real antenna elements and the mirror antenna elements with an HIS as the ground can cancel the mutual coupling between the real antenna elements. It is obvious that the reflection phase of the $\textrm{TM}_{-1,0}$ Floquet mode from the HIS is very critical for large scanning angles.

\begin{figure}[!t]
	\centerline{\includegraphics[width=\columnwidth]{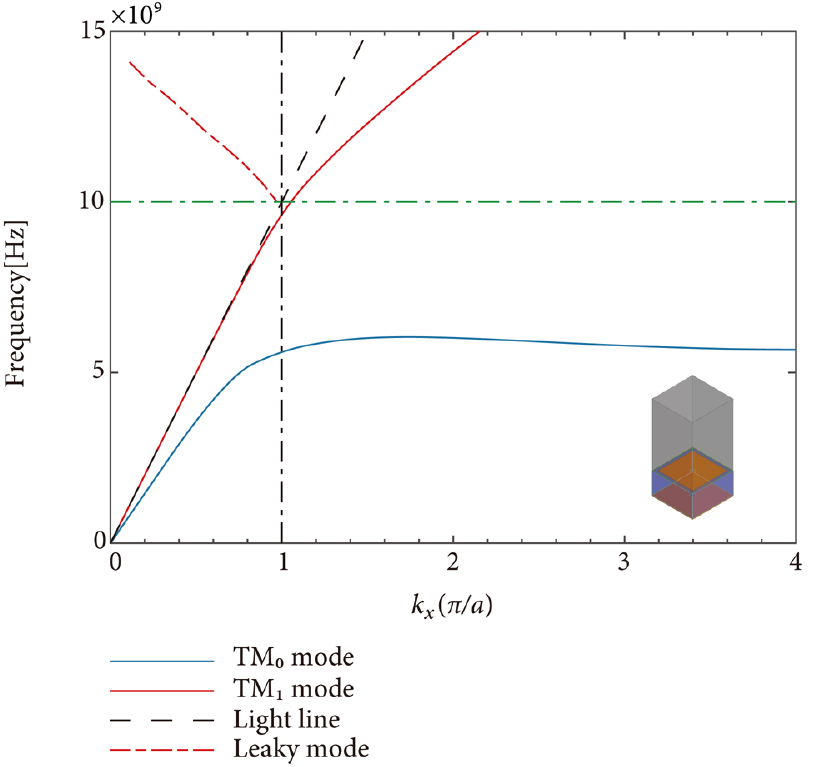}}
	\caption{Dispersion curves of the grounded HIS substrate, where the blue and the red solid lines are the first two SW modes of the HIS; the red dashed line is the LW mode supported by the structure composed of the HIS and the dipole array. The inset is a schematic of a unit cell of the HIS substrate.}
	\label{eigensolve}
\end{figure}

The effects of the reflection phase of the $\textrm{TM}_{-1,0}$ Floquet mode imply the new mechanism which influences the radiating properties of the antenna array. In the next paragraphs, we will reveal the mechanism step by step. First, we show that at large scanning angles this $\textrm{TM}_{-1,0}$ Floquet mode can excite the SW mode supported by the HIS substrate (composed of the HIS ground plane and the dielectric layer above it, which is shown by the inset in Fig. \ref{eigensolve}). Then, we illustrate that this SW mode can be converted into the LW mode by the periodic modulation of the array. When the $\textrm{TM}_{-1,0}$ resonantly excites the LW mode, the LW mode radiation is almost in phase with the direct radiating field from the array, so that the array performance at a large scanning angle could be excellent.

Using eigenmode solver of HFSS, we can calculate the dispersion curves of the unit cell of the HIS substrate. The simulation model of the unit cell is shown as the inset in Fig.\ref{eigensolve}. In the simulation setup, periodic boundaries are used at the sides of the unit cell, and an absorbing material (PML) terminates the setup from the top. The radiating modes from the structure surface propagate within air, filled between the unit cell surface and PML, and are absorbed from the top. The two modes ${\textrm{TM}}_0$ and ${\textrm{TM}}_1$, supported by this HIS substrate are shown in Fig. \ref{eigensolve} by the solid blue and red lines. Since they are lower than the light line which is shown by a black dashed line, both of them are SW modes. Generally, the condition for the existence of such SW modes is $ \theta_{r} + \theta_{up} + 2 k_z d = m \times 2 \pi $, where $ \theta_{r}$ is the reflection phase of the HIS ground, $\theta_{up}$ is the phase of total reflection at the interface between the medium and air, $k_z$ is the wavevector in the $z$ direction in the substrate dielectric material, and $m$ is the order of the SW modes. Clearly, the properties of SW modes are also influenced by the reflection phase of the HIS in a subtle way. In this paper, the working frequency is 10 GHz and the mode of ${\textrm{TM}}_0$ is very far away from this frequency, we neglect the ${\textrm{TM}}_0$ mode in this research. Once the propagation constant of the Floquet mode of antenna array is equal to the propagation constant of ${\textrm{TM}}_1$ mode of HIS substrate, ${\textrm{TM}}_1$ mode will be excited\cite{b11}.

\begin{figure}[!t]
	\centerline{\includegraphics[width=\columnwidth]{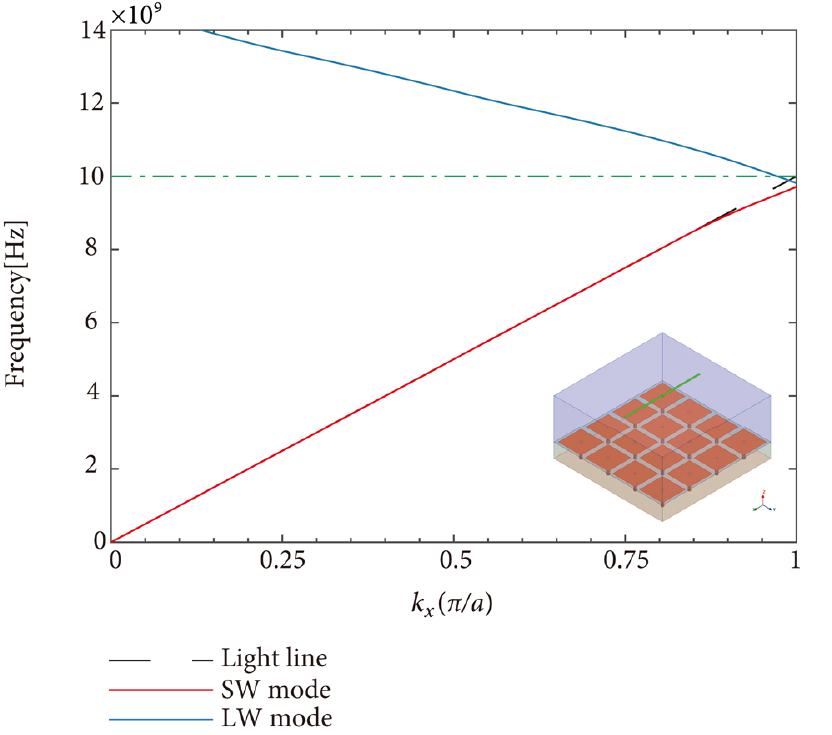}}
	\caption{Dispersion curves of the HIS-based dipole array unit cell using an HFSS simulation. The red and blue solid lines are the SW mode and LW mode of the unit cell respectively. The inset is a schematic of a unit cell of our dipole array.}
	\label{leaky_wave}
\end{figure}

\begin{figure}[!t]
	\centerline{\includegraphics[width=\columnwidth]{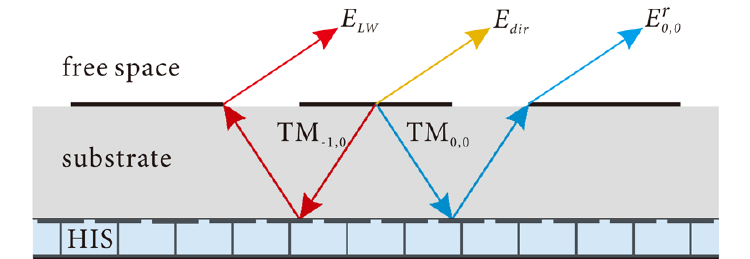}}
	\caption{Schematic illustration of the propagation paths of $E_{dir}$, $E_{0,0}^r$ and $E_{LW}$, where the orange, blue and red lines in the free space are the propagation paths of $E_{dir}$, $E^{r}_{0,0}$ and $E_{LW}$ respectively, while the blue and red lines in the substrate are the propagation paths of $\textrm{TM}_{0,0} $ and $\textrm{TM}_{-1,0}$. }
	\label{dirres}
\end{figure}

However, this ${\textrm{TM}}_1$ mode can be transfer into LW mode. Since the period $a$ of the antenna array is four times larger than the period of the HIS substrate, the dispersion curve of ${\textrm{TM}}_1$ should be folded back if using $a$ as the period, as shown by the red dashed line in Fig. \ref{eigensolve}. Now the red dashed line is above the light line (the black dashed line), which means the SW mode becomes the LW mode which could radiate. Actually, we have also calculated the dispersion curves of the unit cell of the antenna array, which is shown as an inset in Fig. \ref{leaky_wave}. As we expected, the dispersion curve of the mode 2 above the light line is like the dashed line in Fig. \ref{eigensolve}. The physical reason for the transformation of the SW mode to the LW mode is shown in Fig. \ref{dirres}. If the SW mode ${\textrm{TM}}_1$ of the HIS substrate could be excited by an external field, the SW mode will experience the periodic scattering by the antenna array and the scattered field could be a radiating field.

With all this preparation, we now can compose all the pieces together to show the mechanism of radiation with the help of the SW mode. When the antenna array scans at large angles, the Floquet mode $\textrm{TM}_{-1,0}$ becomes the propagating field and it can excite the LW mode, which is from the SW of the HIS substrate with the periodic scattering of the array. Then, with the help of the LW mode, the total radiating performance of the antenna array could be greatly improved for large scanning angles.

Finally, we qualitatively analyze the effect of the LW mode on the scanning performance. As shown in Fig. \ref{dirres}, we decompose the radiation field of the array in free space into three parts: the direct radiation field $E_{dir}$ of the antenna array, the reflected field $E^r_{0,0}$ of the $\textrm{TM}_{0,0}$ mode and the radiation field $E_{LW}$ of the LW mode. Based on the observation that the LW mode field $E_{LW}$ is excited by the Floquet mode $\textrm{TM}_{-1,0}$ in the substrate, the LW mode field has the following general form:
\begin{equation}
E_{LW} = a\frac{\gamma}{j(\omega-\omega_0)+\gamma}
\end{equation}
where $a$, $\omega_0$ and $\gamma$ are the complex amplitude, eigenfrequency and attenuation constant of the LW mode, respectively, and $\omega$ is the operating frequency of the antenna array. Then, the total radiation field $E_{total}$ can be expressed as the sum of the direct radiation field $E_{dir}$, the reflected field $E^r_{0,0}$ of the $\textrm{TM}_{0,0}$ mode and the radiation field from the LW mode excited by $\textrm{TM}_{-1,0}$:
\begin{equation}
E_{total} = E_{dir} + E^r_{0,0} + a\frac{\gamma}{j(\omega-\omega_0)+\gamma}
\end{equation}
When the scanning angle is relatively small, $E_{dir}$ and $ E^r_{0,0}$ dominate the radiation field while the LW mode is difficult to excite and its contribution could be neglected.
As we have discussed in Sec. \ref{mirror}, the special changing of the HIS reflection phase for the $ E^r_{0,0}$ mode can improve the scanning performance. When the scanning angle increases to a value larger than 40$^\circ$, two conditions for the excitation of the LW mode are satisfied. The first condition is that the Floquet mode $\textrm{TM}_{-1,0}$ becomes a propagating wave in the substrate and the second condition is that the LW mode eigenfrequency $\omega_0$ gradually decreases and is close to the antenna working frequency $\omega$. Thus, $E_{LW}$ is almost resonantly excited by $\textrm{TM}_{-1,0}$, and it is nearly in phase with $E_{dir}$. This mechanism explains why the LW mode can help the radiation of this antenna array. Actually, the $E_{LW}$ strengthens with increasing scanning angle. At the angle range from 40$^\circ-$65$^\circ$, the LW mode excitation can help the antenna radiation. However, when the scanning angle becomes very large, e.g., larger than 65$^\circ$, imaginary part of the active impedance rapidly grows, which means that it can absorb most of the energy radiating from the antenna, as shown in Fig. \ref{active_impe}. Finally, the LW excitation can generate scan blindness at approximately 76$^\circ$.

\begin{figure}[!t]
	\subfigure[]{
		\label{fig:leakywave:a} 
		\begin{minipage}[b]{0.5\columnwidth}
			\centering
			\includegraphics[scale=0.9]{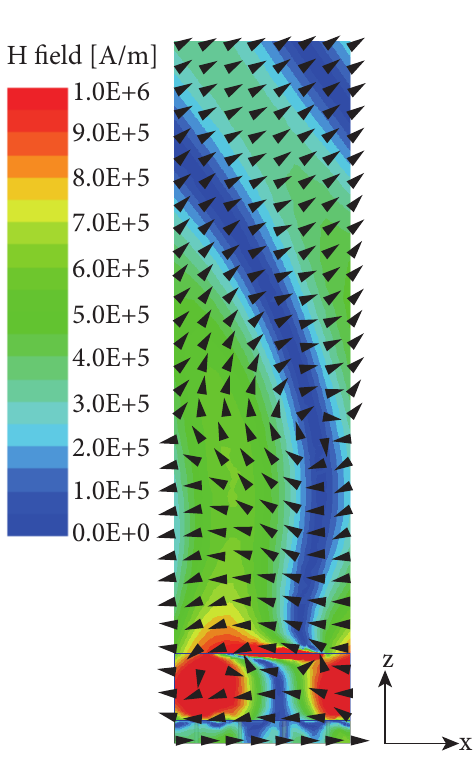}
	\end{minipage}}%
	\subfigure[]{
		\label{fig:scanblindness:b} 
		\begin{minipage}[b]{0.5\columnwidth}
			\centering
			\includegraphics[scale=0.9]{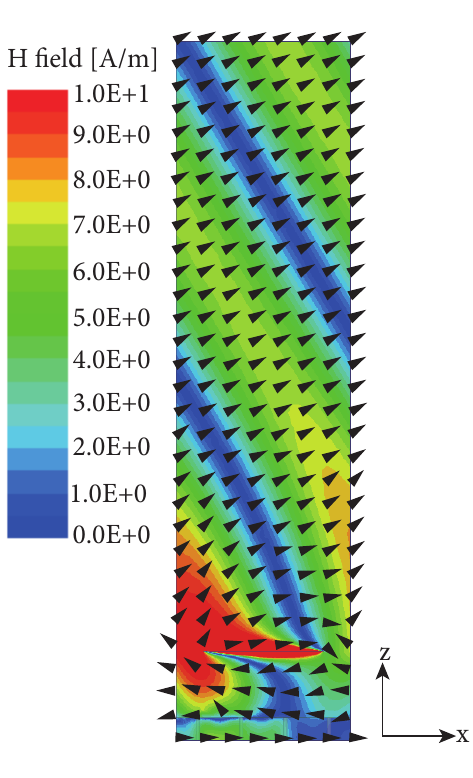}
	\end{minipage}}
	\caption{Comparison of field distributions and energy flow directions between the LW eigenfield and the radiation field of the array at a radiation angle of $60^{\circ}$, where the color distribution and the arrow direction represent the magnitude of the magnetic field and the direction of the Poynting vector, respectively. (a) LW mode eigenfield and the direction of the Poynting vector, (b) total field and the direction of the Poynting vector of our array.}
	\label{leaky_wave_comp} 
\end{figure}

To demonstrate that the LW mode is truly excited in our model, we have shown the field and the Poynting vector distribution of the LW mode in Fig. \ref{fig:leakywave:a} and the total field of our antenna array at a scanning angle of 60$^\circ$ in Fig. \ref{fig:scanblindness:b}. As we predicted in Fig. \ref{dirres}, when the LW mode is excited by the $\textrm{TM}_{-1,0}$ mode, the Poynting vector should be in the direction opposite to that of the radiation in the substrate, which is true in Fig. \ref{fig:leakywave:a} and \ref{fig:scanblindness:b}. From Fig. \ref{fig:scanblindness:b}, we also can see that the direct of the radiation field on the antenna surface is nearly in phase with the LW mode since both are shown in red.

\section{Conclusion}
\label{conclusion}
An infinite two-dimensional dipole array with a mushroom-like HIS ground plane is designed, which can achieve a wide scanning angle of up to $65^\circ$ in the elevation plane. The active impedance and $S11$ of the array calculated via theoretical Floquet analysis are in good agreement with numerical simulation results. Two new mechanisms which support the excellent performance of such an array at a wide scanning angle are demonstrated theoretically and numerically. In the range of small scanning angles, these excellent properties are mainly from the special reflection phase of the HIS ground, which can cause the mutual coupling between the elements of a real array be compensated by the mutual coupling effect from the mirror array. For the range of large scanning angles, since the surface wave (SW) mode could be resonantly excited by high order Floquet mode $\textrm{TM}_{-1,0}$ from the array and the SW mode could be converted into a leaky wave mode by the scattering of the array, the radiation field from the LW mode is nearly in phase with the direct radiating field from the array. Therefore, with the help from the special reflection phase of the HIS and the designed LW mode on the HIS ground, the antenna array with an HIS ground can achieve wide-angle scanning performance. We think these mechanisms could be widely used in the design of wide-angle scanning arrays.

\section*{Data Availability}
The data used to support the findings of this study are available from the corresponding author upon request.

\section*{Conflicts of Interest}
The authors declare that there are no conflicts of interest regarding the publication of this paper.


\begin{thebibliography}{00}
	
	\bibitem{b11} D. M. Pozar, Daniel H. S., ``Scan blindness in infinite phased arrays of printed dipoles,'' \emph{IEEE Transactions on Antennas and Propagation}, vol. 32, no. 6, pp. 602-610, 1984.
	
	\bibitem{b1} Qamar, Z., Riaz, L., Chongcheawchamnan, M., Khan, S. A., \& Shafique, M. F.  ``Slot combined complementary split ring resonators for mutual coupling suppression in microstrip phased arrays,'' \emph{IET Microwaves, Antennas \& Propagation}, vol. 8, no. 15, pp. 1261-1267, 2014.
	
	\bibitem{b2} Tsay, W-J and Pozar, David M, ``Radiation and scattering from infinite periodic printed antennas with inhomogeneous media,'' \emph{IEEE Transactions on Antennas and Propagation},  vol. 46, no. 11, pp. 1641-1650, 1998.
	
	\bibitem{b3} Yazdi, Shirin Ramezanzadeh, Somayye Chamaani, and Seyed Arash Ahmadi ``Mutual Coupling Reduction in Microstrip Phased Array Using Stacked-Patch Reduced Surface Wave Antenna,'' \emph{Antennas and Propagation \& USNC/URSI National Radio Science Meeting, 2015 IEEE International Symposium on} (pp. 436-437). IEEE.
	
	\bibitem{b4} Moghadas, H., A. Tavakoh, M. Salehi, ``Elimination of scan blindness in microstrip scanning array antennas using defected ground structure,'' \emph{AEU-International Journal of Electronics and Communications},  vol. 62, no. 2, pp. 155-158, 2008.
	
	\bibitem{b5} Sievenpiper, Dan \emph{et al., ``}High-impedance electromagnetic surfaces with a forbidden frequency band,'' \emph{IEEE Transactions on Microwave Theory and techniques},  vol. 47, no. 311, pp. 2059-2074, 1999.
	
	\bibitem{b6} Adas, Enver, Franco De Flaviis, and Nicolaos G. Alexopoulos. ``Integrated Microstrip Antennas and Phased
	Arrays with Mode-Free Electromagnetic Bandgap Materials for Scan Blindness Elimination'' \emph{Electromagnetics},  vol. 37, no. 1, pp. 2000-2007, 2004.
	
	\bibitem{Donzelli2007} Donzelli, G., Capolino, F., Boscolo, S., and Midrio, M. ``Elimination of scan blindness in phased array antennas using a grounded-dielectric EBG material.'' \emph{IEEE Antennas and Wireless Propagation Letters}, vol 6, pp. 106-109, 2007.
	
	\bibitem{Li2007} Li, L., C-H. Liang, and C-H. Chan. ``Waveguide end-slot phased array antenna integrated with electromagnetic bandgap structures.'' \emph{Journal of Electromagnetic Waves and Applications}, vol 21, no. 2, pp. 161-174, 2007.
	
	\bibitem{b7} Fu Y Q, Yuan N C, ``Elimination of scan blindness in phased array of microstrip patches using electromagnetic bandgap materials,'' \emph{IEEE Antennas and Wireless Propagation Letters},  vol. 3, no. 1, pp. 63-65, 2004.
	
	\bibitem{Azarbar2011} Azarbar, A., and J. Ghalibafan. ``A compact low-permittivity dual-layer EBG structure for mutual coupling reduction.'' \emph{International Journal of Antennas and Propagation}, vol. 2011, 2011.
	
	\bibitem{b8} Costa F \emph{et al., ``}TE surface wave resonances on high-impedance surface based antennas: Analysis and modeling,'' \emph{IEEE Transactions on Antennas and Propagation},  vol. 59, no. 10, pp. 3588-3596, 2011.
	
	\bibitem{b9} Li M \emph{et al., ``}Compact surface-wave assisted beam-steerable antenna based on HIS,'' \emph{IEEE Transactions on Antennas and Propagation},  vol. 62, no. 7, pp. 3511--3519, 2014.
	
	\bibitem{b10} Li M, Xiao S Q, Wang B Z ``Investigation of using high impedance surfaces for wide-angle scanning arrays,'' \emph{IEEE Transactions on Antennas and Propagation}, vol. 63, no. 7, pp. 2895-2901, 2015.
	
	\bibitem{Tamir1973} Tamir, T. ``Inhomogeneous wave types at planar interfaces: III-Leaky waves.'' \emph{Optik}, vol. 38,  pp. 269-297, 1973.
	
	\bibitem{bhattacharyya2006phased} \emph{Phased array antennas: Floquet analysis, synthesis, BFNs and active array systems}, Bhattacharyya Arun K., USA: John Wiley \& Sons, 2006.
\end{thebibliography}
\end{document}